\documentclass[aps,prl,twocolumn,showpacs,showkey,square,numbers,amssymb,amsmath,nobibnotes]{revtex4-1}
\usepackage{bm}
\usepackage{times}
\usepackage{graphicx}
\usepackage[usenames,dvipsnames,svgnames,table]{xcolor}
\usepackage{hyperref}
\hypersetup{colorlinks=true, linkcolor=NavyBlue, citecolor=PineGreen,urlcolor=cyan}

\newcommand{\bH     }{\mbox{\boldmath$H$}}

\newcommand{\bracket}[1]{\left\langle #1\right\rangle}

\begin{document}
\title{Large deviation function for the number of eigenvalues of sparse random graphs inside an interval}
\author{Fernando L. Metz}
\address{Departamento de F\'isica, Universidade Federal de Santa Maria, 97105-900 Santa Maria, Brazil}
\author{Isaac P\'erez Castillo}
\address{Department of Complex Systems, Institute of Physics, UNAM, P.O. Box 20-364, 01000 M\'exico, D.F., M\'exico}
\begin{abstract}
We present a  general method to obtain the exact rate function $\Psi_{[a,b]}(k)$ controlling the large deviation probability  $\text{Prob}[\mathcal{I}_N[a,b]=kN]  \asymp e^{-N\Psi_{[a,b]}(k)}$  that an $N \times N$  sparse random matrix has $\mathcal{I}_N[a,b]=kN$ eigenvalues inside the interval $[a,b]$. The method is applied to study the eigenvalue statistics in two distinct examples: (i) the shifted index number of eigenvalues for an ensemble of Erd\"os-R\'enyi graphs and (ii) the number of eigenvalues within a bounded region of the spectrum for the Anderson model on regular random graphs. A salient feature of the rate function in both cases is that, unlike rotationally invariant random matrices, it is asymmetric with respect to its minimum. The asymmetric character depends on the disorder in a way that is compatible with the distinct eigenvalue statistics corresponding to localized and delocalized eigenstates. The results also show that the level compressibility $\kappa_2/\kappa_1$ for the Anderson model on a regular graph fulfills $0 < \kappa_2/\kappa_1 < 1$ in the bulk regime, in contrast to the behavior found in Gaussian random matrices.
Our theoretical findings are thoroughly compared to numerical diagonalization in both cases, showing a reasonable good agreement.
\end{abstract}
\maketitle

Since the fundamental works of Wigner \cite{Wigner} and Dyson \cite{Dyson1,Dyson2,Dyson3} that lay the foundations of random matrix theory, several observables related to the eigenvalue statistics of $N \times N$ random matrices have been studied and a wealth of quantitative information is currently available, constituting an invaluable tool to address problems in various disciplines \cite{BookMehta}. 
A primary observable is the number of eigenvalues $\mathcal{I}_N[a,b]$ within a {\it bounded} interval $[a,b]$ on the real line. The statistics of $\mathcal{I}_N[a,b]$ describes the ground-state fluctuations of many-body systems \cite{Marino2014,Marino2016}, whose experimental realization may be achieved by confining cold atoms in optical laser traps \cite{Bloch2008}. From a more theoretical perspective, the fluctuations of $\mathcal{I}_N[a,b]$ provide a criterion to distinguish between the localized and the extended phase in non-interacting disordered electronic systems \cite{BookMehta,Altshuler88}, due to the striking different behavior of the eigenvalue statistics in each phase.  Several works have been also devoted to the statistics of the number of eigenvalues in an {\it unbounded} interval $(-\infty,b]$ \cite{Cavagna2000,Majumdar1,Majumdar2,Vivo,Majumdar3,Perez2014b, Perez2014c,Perez2015, Texier2016}, which is relevant to problems in different areas, such as the study of the intricate energy landscape of disordered systems \cite{Broderix2000,Cavagna2000,BookWales}, or the meaningful analysis of the correlation matrix built from large datasets \cite{Vivo,Perez2015}. These works deal with rotationally invariant random matrices (RIRM), where the joint distribution of eigenvalues is analytically known and the Coulomb gas method can be applied, yielding analytical results not only for {\it typical} statistical fluctuations of $\mathcal{I}_N[a,b]$, but also for {\it atypical}, rare fluctuations, which remain finite for $N \rightarrow \infty$.

Although the Coulomb gas method has played a crucial role in random matrix theory, its application is limited to RIRM.  The statistics of $\mathcal{I}_N[a,b]$ in other interesting random matrix ensembles has eluded a careful treatment, as the analytical form of the joint distribution of eigenvalues is not generally known. In this sense, the most relevant examples come from spectral graph theory \cite{BookGraphSpectra}, in which the central interest lies in the eigenvalue statistics of certain matrices related to sparse random graphs, defined as a set of $N$ nodes connected randomly by edges. The behavior of the fluctuations of $\mathcal{I}_N[a,b]$ in random graphs is an interesting subject from the theoretical side, due to the interplay between the distinct statistical properties of eigenvalues corresponding to localized and extended states, both usually coexisting in the spectra of random graphs \cite{Evangelou1,Evangelou2,Metz2010,Biroli2010,Slanina2012}. In the last decade, random graphs have become a fundamental tool to explore different branches of science, finding applications in complex networks, spin-glasses and information theory (see \cite{MezardBook,BarratBook} and references therein). Another important application is the study of transport properties in disordered electronic systems, where random graphs give rise to mean-field models \cite{Abou73,Evangelou1,Evangelou2,Fyod,Biroli2010,Biroli2012}. Motivated by the connection between Anderson localization on a regular random graph (RRG) and localization in the Fock space of many-body quantum systems \cite{Altshuler1997,Basko2006}, there has been a renewed interest  in the Anderson model on a RRG  due to the possible existence of a novel, non-ergodic delocalized phase \cite{Biroli2012,Scar2014,Scar2014a,kravtsov2015,Shukla2016,Li2016,Mirlin2016,Alt2016}, which would be characterized by extended eigenstates corresponding to uncorrelated energy levels \cite{Biroli2012}. In spite of this ubiquitousness, analytical techniques to pursue an in-depth analysis of the eigenvalue fluctuations of random graphs are still lacking, even in the context of the well-studied Anderson model on a RRG. 

In this paper we introduce a powerful method to compute analytically the rate function $\Psi_{[a,b]}(k)$ describing the large deviations that a large $N \times N$ matrix associated to a random graph model has $\mathcal{I}_N[a,b]=kN$ eigenvalues inside $[a,b]$. Our approach explores an analogy between spin-glasses and random matrices by mapping the problem of computing the cumulant generating function (CGF) of $\mathcal{I}_N[a,b]$ in a free-energy calculation reminiscent from spin-glasses, which can be pursued using the replica method \cite{BookParisi}.  In order to illustrate the general character of our technique, we present results for two different examples: (i) the rate function of $\mathcal{I}_N(-\infty,L]$ for Erd\"os-R\'enyi (ER) random graphs; (ii) the statistics of $\mathcal{I}_N[-L,L]$ for the Anderson model on a RRG. As a common finding, the rate function of $\mathcal{I}_N$ is asymmetric with respect to its minimum, in contrast to its symmetric nature for RIRM studied up to the present. We argue that such asymmetry comes from the presence of both localized and extended states in the spectra of random graphs.  As another outcome of the method, our results show that, for fixed $L=O(1)$ and large $N$, the level compressibility  $\kappa_2/\kappa_1$  \cite{Chalker1996,Bogomolny2011} for the Anderson model on a RRG fulfills $0 < \kappa_2/\kappa_1 < 1$, which complies with the absence of strong level repulsion. All results are compared with numerical diagonalization of large random matrices, showing a fairly good agreement.

We consider an $N\times N$ symmetric real matrix $\bH$ with eigenvalues $\lambda_1,\ldots,\lambda_N$, where $\mathcal{I}_N[a,b]$ denotes the number of eigenvalues inside $[a,b]\subseteq\mathbb{R}$. If $\rho_N(\lambda)=(1/N)\sum_{i=1}^N\delta(\lambda-\lambda_i)$, obviously
\begin{equation}
 \begin{split}
\mathcal{I}_N[a,b]= N\int_{a}^b d\lambda \, \rho_N(\lambda)\,.
\end{split}
\end{equation}
Using the discontinuity of the complex logarithm along the branch cut on the negative real axis as a prescription
of the Heaviside step function $\Theta(-x)=\frac{1}{2\pi i}\lim_{\eta\to 0^{+}}[\ln (x+i\eta)-\ln (x-i\eta)]$, one derives the expression
 \begin{equation}
 \begin{split}
\mathcal{I}_N[a,b]&=-\frac{1}{\pi i}\lim_{\eta\to0^{+}}\ln\left[\frac{Z(b_{\eta})Z({a}^\star_{\eta})}{Z(b^\star_\eta)Z(a_{\eta})}\right]
\label{eq:pin}\,,
\end{split}
\end{equation}
where $a_{\eta}\equiv a-i\eta$ and  $b_{\eta}\equiv b-i\eta$. We have introduced $Z(z)=\left[\det\left(\bH-z\openone\right)\right]^{-1/2}$, with   $\openone$ the $N\times N$ identity matrix and $(\cdots)^\star$ the complex conjugation. Next we introduce the CGF for the statistics of $\mathcal{I}_N[a,b]$ 
 \begin{equation}
 \begin{split}
\mathcal{F}_{[a,b]}(y)=-\lim_{N\to\infty}\frac{1}{N}\ln \bracket{e^{-y\mathcal{I}_N[a,b] }} \,,
\label{eq:cgf1}
\end{split}
\end{equation}
where $\bracket{\cdots}$ represents the average over the ensemble of random matrices $\bH$, specified through the distribution $p(\bH)$. Combining eqs. \eqref{eq:pin} and \eqref{eq:cgf1}, one can write 
\begin{equation}
 \begin{split}
\mathcal{F}_{[a,b]}(y)=-\lim_{N\to\infty} \lim_{\eta\to 0^+}\frac{1}{N}\ln \mathcal{Q}_{[a_\eta,b_\eta]}(y)\,,
\label{sda}
\end{split}
\end{equation}
with
\begin{equation}
 \begin{split}
\mathcal{Q}_{[a_\eta,b_\eta]}(y )=\bracket{ Z^{\frac{i y}{\pi }}({b}^\star_\eta) Z^{\frac{i y}{\pi }}(a_{\eta})Z^{-\frac{i y}{\pi }}(b_\eta)Z^{-\frac{i y}{\pi }}({a}^\star_{\eta}) } \,.
\label{eq:fq}
\end{split}
\end{equation}
Assuming that $\mathcal{F}_{[a,b]}(y)$ is differentiable for arbitrary $y \in \mathbb{R}$ \cite{Touchette}, from large
deviation theory we have that $\text{Prob}[\mathcal{I}_N[a,b]=kN] \asymp e^{-N\Psi_{[a,b]}(k)}$, where the rate function $\Psi_{[a,b]}(k)$ \cite{BookDembo}
is related to the CGF $\mathcal{F}_{[a,b]}(y)$ by the Legendre transform
\begin{equation}
 \begin{split}
-\Psi_{[a,b]}(k)=\underset{y\in \mathbb{R}}{\text{inf}}\left[ k y-\mathcal{F}_{[a,b]}(y)\right]\,,
\label{eq:rf}
\end{split}
\end{equation}
while the $\ell$th cumulant $\kappa_\ell[a,b]$ of $\mathcal{I}_N[a,b]$ follows from
\begin{equation}
 \begin{split}
\kappa_\ell[a,b]=(-1)^{\ell+1}\frac{\partial ^\ell \mathcal{F}_{[a,b]}(y)}{\partial y^\ell}\Big|_{y=0}\,.
\label{fgh}
\end{split}
\end{equation}
Thus, the CGF is the central object of interest, since the computation of $\Psi_{[a,b]}(k)$ and $\kappa_\ell[a,b]$ boils down to being able to determine $\mathcal{F}_{[a,b]}(y)$.

Fortunately, $\mathcal{F}_{[a,b]}(y)$ can be calculated exactly for $N \rightarrow \infty$ using spin-glass techniques \cite{BookParisi}. According to eqs. \eqref{sda} and \eqref{eq:fq}, $\mathcal{F}_{[a,b]}(y)$ is obtained from the ensemble average of imaginary powers of $Z(z)$, which is an unworkable calculation. In order to overcome this obstacle, one employs the replica method as discussed in \cite{Metz2015}, by defining the function
\begin{equation}
 \begin{split}
\mathcal{Q}_{r}(n_{+},n_{-})= \Big{\langle} \left[Z({b}^\star_\eta) Z(a_{\eta}) \right]^{n_{+}} \left[ Z(b_\eta)Z({a}^\star_{\eta}) \right]^{n_{-}} \Big{\rangle}  \,
\end{split}
\end{equation}
in terms of positive integers $n_{\pm}$. Once $\mathcal{Q}_{r}(n_{+},n_{-})$ is computed in the limit $N \rightarrow \infty$, the function $\mathcal{Q}_{[a_\eta,b_\eta]}(y)$ of Eq. (\ref{eq:fq}) is recovered by making an analytical continuation of $n_{\pm}$ to the complex plane and then performing the replica limit $n_{\pm} \rightarrow \pm i y / \pi$  of $\mathcal{Q}_{r}(n_{+},n_{-})$. 
Although the general scheme of the replica approach, including the underlying interchange  of limits $N \rightarrow \infty$ and $n_{\pm} \rightarrow \pm i y / \pi$, has been rigorously established only for some disordered systems \cite{Talagrand}, the replica method has proven to be a valuable tool to calculate exactly the averaged spectral properties of random
matrices for $N \rightarrow \infty$ (see \cite{Note1} and references therein). All technical details of the replica method to compute $\mathcal{Q}_{[a_\eta,b_\eta]}(y)$ are discussed in the supplemental material \footnote{See Supplemental Material, which includes Refs. \cite{Dean2002,Ergun,Rogers1,KuhnSmall,Bai,MezPar,MezPar1}.}.

In order to illustrate the versatility  of our approach, we study two different examples: (i) the number of eigenvalues inside $(-\infty,L]$, also known as the shifted index number (SIN), for the adjacency matrix of Erd\"os-R\'enyi (ER) graphs \cite{Bollobas}; (ii) the number of eigenvalues within $[-L,L]$ for the Hamiltonian describing the Anderson model on a RRG \cite{Wormald}. The statistics of $\mathcal{I}_N$ is studied from the eigenvalues of an $N \times N$ symmetric random matrix $\bH$. Both models are defined by a common matrix $\bH$ with entries $H_{ij} = \epsilon_i \delta_{ij} + c_{ij}$, where $\{ \epsilon_i \}$  are independent random variables drawn from the distribution $P_\epsilon(\epsilon)$. 

The structure of each random graph is encoded in the entries $\{ c_{ij} \}$ of the underlying adjacency 
matrix \cite{Bollobas}: $c_{ij} = 1$ if nodes $i$ and $j$ are connected, and $c_{ij} = 0$ otherwise. 
The distributions of $\{ c_{ij} \}$ for each example are presented in \cite{Note1}. It is important to note that, for a RRG, the number
of neighbors connected to each node is fixed to a integer $c$, while this quantity fluctuates from node to node in the case
of ER random graphs, with an average value $c \in \mathbb{R}$. We refer to \cite{Note1} for further details regarding the definition of each random graph model. 

We present below the main outcomes of the method, namely the analytical results for the rate functions in each case.
Let  $\mathcal{F}_{L}^{(p)}(y)$ and $\mathcal{F}_{L}^{(a)}(y)$ denote, respectively, the CGF's for the examples (i) and (ii) introduced above. 
After following the replica method  \cite{Note1}, one ends up with the expressions 
\begin{align}
\mathcal{F}_{L}^{(p)}(y)&= \frac{c}{2} \int du \, dv \, \omega_p (u) \omega_p(v) \left[ 
e^{\frac{y}{\pi} \varphi(u,v) }  \nonumber - 1 \right]  \\
& - \ln{\left[\int du \, \mu (u) e^{\frac{y}{\pi} \theta(u)}\right]} \,,
\label{eq:cgfP}  \\
\mathcal{F}_{L}^{(a)}(y) &=  \frac{1}{2} (c-2) \ln\left[ \int du \, dv~ \nu(u,v|c) e^{\frac{ y}{\pi} \left[ \theta(u) + \theta(v)   \right]} \right] \nonumber \\
&  - \frac{c}{2} \ln\left[ \int du \, dv ~\nu(u,v|c-1) e^{ \frac{ y}{\pi} \left[ \theta(u) + \theta(v)   \right]  } \right]  \,,\label{jsoi} 
\end{align}
where $u$ and $v$ are complex variables, and we have defined 
\begin{equation}
 \begin{split}
\hspace{-.1cm}\theta(u) &= - \frac{i}{2} \ln \left(\frac{u}{u^{\star}}\right)\,,~ \varphi(u,v) = - \frac{i}{2} \ln \left[ \frac{1 + \frac{1}{u v}}{1 + \frac{1}{(u v)^{\star}}}  \right]\,.
\end{split}
\end{equation}
The integrals with the measure $du \,  dv $ in Eqs. (\ref{eq:cgfP}) and (\ref{jsoi}) run
over all possible values of the real and imaginary parts of $u$ and $v$, with the constraints
${\rm Re} \, u > 0$ and  ${\rm Re} \, v > 0$ \cite{Note1}.

The function $\mu(u)$ is the joint distribution of $({\rm Re} \, u,{\rm Im} \, u)$, while $\nu(u,v|c)$ is the joint
distribution of the real and imaginary parts of $u$ and $v$ for a fixed $c$. These quantities are evaluated from
\begin{align}
\mu(u) &= \sum_{k=0}^{\infty} \frac{e^{-c} c^{k}}{k!} \int\left[\prod_{n=1}^k  d u_n \omega_p(u_n)\right]  \nonumber \\
&\times \Big\langle \delta\left[ u - F_\epsilon(u_{1,\dots,k})  \right] \Big\rangle_{\epsilon} \,, \label{ddf} \\
\nu(u,v|c) &= \int \left[ \prod_{n=1}^{c} du_n \,  dv_n \, \omega_a (u_n,v_n)  \right] \nonumber \\
&\times \Bigg\langle\delta\left[ u - F_\epsilon(u_{1,\dots,c})  \right] \delta\left[ v   - F_{-\epsilon}(v_{1,\dots,c})    \right]\Bigg\rangle_{\epsilon}\,, \label{ddfa}
\end{align}
with 
\begin{equation}
 \begin{split}
F_\epsilon(u_{1,\dots,k}) = i \left( \epsilon - z^{*}   \right) + \sum_{n=1}^{k} \frac{1}{u_n}\, ,  
\end{split}
\end{equation}
and $z = L - i \eta$. The symbol $\langle \dots \rangle_{\epsilon}$ denotes the average 
over $\epsilon$. The system of equations is closed for each example by the equations for the joint distributions $\omega_p(u)$ and $\omega_a (u,v)$
\begin{align}
\omega_p(u) &= \frac{e^{\frac{y}{\pi} \theta(u)} \mu(u)}{ \int du \, e^{\frac{y}{\pi} \theta(u)} \mu(u) } \, , \label{eq:sce} \\
\omega_a (u,v) &= \frac{e^{ \frac{ y}{\pi} \left[ \theta(u) + \theta(v)   \right]  } \nu(u,v|c-1)}{\int du \, dv \,  e^{ \frac{ y}{\pi} \left[ \theta(u) + \theta(v)   \right]  } \nu(u,v|c-1) }\,. \label{eq:sce1}
\end{align}
The limit $\eta \rightarrow 0^{+}$ is implicit in Eqs. (\ref{eq:cgfP}) and (\ref{jsoi}) as well as in the equations for the
distributions. 

The system of Eqs. (\ref{ddf}-\ref{eq:sce1}) determine all distributions needed to calculate $\mathcal{F}_{L}^{(p)}(y)$ and $\mathcal{F}_{L}^{(a)}(y)$.
By substituting Eqs. (\ref{ddf}) and (\ref{ddfa}) in Eqs. \eqref{eq:sce} and \eqref{eq:sce1}, we obtain self-consistent equations for $\omega_p(u)$ and $\omega_a (u,v)$, whose solutions depend on $y$. As the $y$-dependent factors play the role of reweighting terms in Eqs. \eqref{eq:sce} and \eqref{eq:sce1}, these are solved numerically by a weighted population dynamics algorithm, discussed in \cite{Note1}. The subsequent numerical solutions are used to evaluate the CGF's of Eqs. \eqref{eq:cgfP} and \eqref{jsoi} for different values of $y$, and the corresponding rate functions $\Psi_{L}^{(p)}(k)$ and $\Psi_{L}^{(a)}(k)$ follow from Eq. \eqref{eq:rf}. 
For $y=0$, Eqs. \eqref{eq:sce} and \eqref{eq:sce1} have a standard form, already found in similar problems \cite{Kuhn2008,Perez2008,Metz2010,Metz2015}.  

\begin{figure}[h!]
\begin{center}
\includegraphics[width=8cm,height=5.5cm]{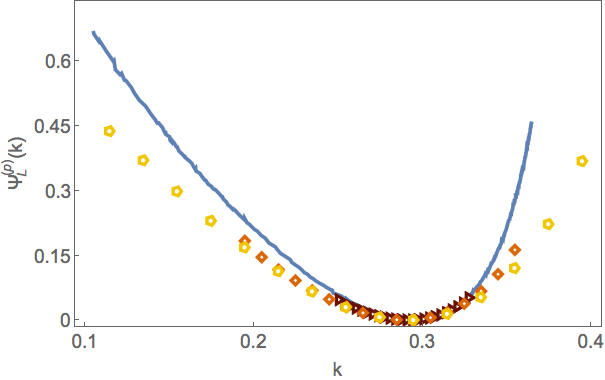}
\caption{Rate function $\Psi_{L}^{(p)}(k)$ of the fraction of eigenvalues inside $(-\infty,L]$ for Erd\"os-R\'enyi graphs with $L=-1$ and average connectivity $c=3$. The solid line is the population dynamics results and the symbols correspond to numerical diagonalization of matrices of sizes $N=50$ (yellow pentagons), $N=100$ (orange rhombic symbols) and $N=300$ (dark-red triangles), using ensembles with $7\times 10^9$, $6 \times 10^8$ and $2\times 10^8$ samples, respectively.}
\label{Poissonfig1}
\end{center}
\end{figure}
Firstly we present results for the rate function $\Psi_{L}^{(p)}(k)$ governing the statistics of $I_N(-\infty,L]$ for ER graphs with $P_{\epsilon}(\epsilon) = \delta(\epsilon)$. The function $\Psi_{L}^{(p)}(k)$ for $c=3$ is displayed in figure \ref{Poissonfig1}, where we compare the population dynamics results with  numerical diagonalization of finite matrices. Since the probability of observing $\mathcal{I}_N(-\infty,L]= k N$ behaves as $e^{-N \Psi_{L}^{(p)}(k)}$ for $N \gg 1$, there is a compromise between considering larger and larger $N$ to suppress finite size effects while at the same time exploring a sizeable subinterval of $k\in[0,1]$. In spite of this difficulty, numerical diagonalization results approach the theoretical results for increasing $N$.

\begin{figure}[h!]
\begin{center}
\includegraphics[width=8cm,height=5.5cm]{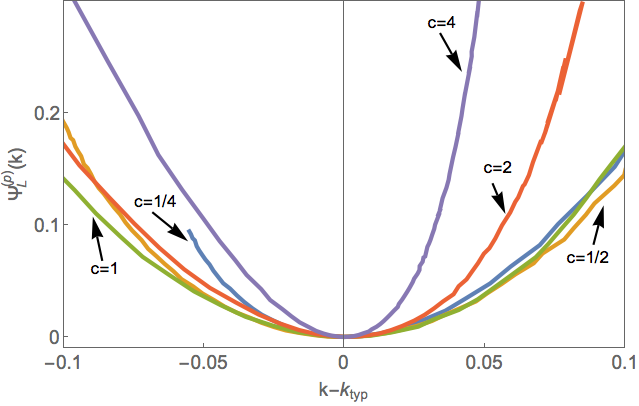}
\caption{Population dynamics results for the rate function $\Psi_{L}^{(p)}(k)$ of the fraction of eigenvalues inside $(-\infty,L]$ for Erd\"os-R\'enyi graphs with $L=-1$ and different values of the average connectivity $c$. The typical value $k_{\rm typ}$ of the shifted index is defined in the main text.}
\label{Poissonfig2}
\end{center}
\end{figure}

The effect of the average connectivity $c$ on $\Psi_{L}^{(p)}(k)$ is illustrated in figure \ref{Poissonfig2}. A notable feature of $\Psi_{L}^{(p)}(k)$ is  its asymmetry around the position of its minimum, located  at the typical value $k_{\text{typ}}=\lim_{N\to\infty}\bracket{\mathcal{I}_N(-\infty,L]}/N$. This is at odds to the behavior of the rate functions describing the eigenvalue statistics  in RIRM studied up to the present \cite{Majumdar1,Majumdar2,Vivo,Majumdar3,Marino2014,Perez2014b,Perez2014c,Perez2015,Texier2016}, but consistent with the gradual change of the eigenvalue statistical properties as $c$ increases \cite{Evangelou1,Evangelou2}. For $c < 1$, the graph is composed of finite, disconnected clusters, and all eigenvectors are localized \cite{Evangelou2,Mirlin91}, while a giant cluster emerges at $c \geq 1$, with the spectrum presenting a mobility edge that separates localized and extended eigenstates \cite{Mirlin91,Metz2010,Slanina2012}. Level repulsion between neighboring eigenvalues is stronger for $c \geq 1$ and, accordingly, samples that increase the SIN become less probable, resulting in rate functions that grow faster for $k > k_{\text{typ}}$ when compared to the left branch $k < k_{\text{typ}}$. By rescaling $c_{ij}$ as $c_{ij} \rightarrow c_{ij}/\sqrt{c}$, $\Psi_{L}^{(p)}(k)$ becomes gradually more symmetric for increasing $c > 1$ \cite{Note1}, consistently with RIRM \cite{Majumdar1,Majumdar2,Vivo,Majumdar3,Marino2014,Perez2014b,Perez2014c,Perez2015,Texier2016}. 

\begin{figure}[h]
\includegraphics[width=8cm,height=5.5cm]{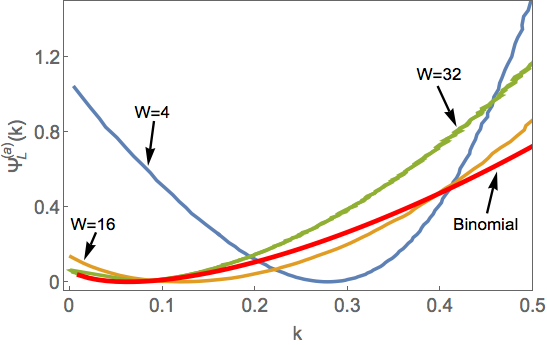}
\caption{Population dynamics results for the rate function $\Psi_L^{(a)}(k)$ of the fraction of eigenvalues inside $[-L,L]$ for the Anderson model on a regular random graph with fixed connectivity $c=3$, $L=1$ and different disorder strengths $W$. The solid red line is the rate function of a binomial distribution, where $2L/W$ is the probability that an eigenvalue falls into $[-L,L]$.}
\label{Andersonfig1}
\end{figure}

Next we present results for the rate function $\Psi_{L}^{(a)}(k)$ controlling the fraction of eigenvalues inside $[-L,L]$ for the Anderson model on a RRG. The diagonal entries $\epsilon_1,\dots,\epsilon_N$ are uniformly distributed in $[-W/2,W/2]$. The statistics of $\mathcal{I}_N[-L,L]$ depends crucially on how $L$ scales with $N$ \cite{Marino2014}. Below we comment on the possibility to apply our method to study the local eigenvalue statistics, obtained by considering $L=O(1/N)$ \cite{BookMehta,Mirlin}. Here we limit ourselves 
to the regime where $L=O(1)$, independently of $N$, such that $\rho(\lambda)$ is not uniform over $[-L,L]$. In this case, the asymmetric nature of $\Psi_{L}^{(a)}(k)$ changes as a function of $W$, similarly to ER graphs, as shown in figure \ref{Andersonfig1}. Repulsion between neighboring levels  becomes more prominent for smaller $W$, which makes the fluctuations that tend to raise $\mathcal{I}_N[-L,L]$ rarer. For $W > W_c \simeq 17.5$, all eigenstates are localized and the level-spacing distribution corresponding to the local eigenvalue statistics 
follows a Poisson law \cite{Abou73,Biroli2010,Biroli2012}, such that the eigenvalues behave as uncorrelated random variables. The rate function is closer to that of a binomial distribution for large $W$, since $\rho(\lambda)$ becomes approximately uniform over $[-L,L]$.

We finish by presenting results for the cumulant ratio $\kappa_2/\kappa_1$ of $\mathcal{I}_{N}[-L,L]$  for the Anderson model on a RRG. From Eq. (\ref{fgh}), we have that $\kappa_2 / \kappa_1 = \sigma_{N}^{2}/m_N$ is the level compressibility \cite{Altshuler88,Chalker1996,Bogomolny2011}, since $\sigma_{N}^{2} = \langle \mathcal{I}_{N}^2 \rangle - \langle \mathcal{I}_{N} \rangle^2$ is the number variance  and $m_N = \langle \mathcal{I}_{N} \rangle$ is the mean number of eigenvalues inside $[-L,L]$. The analytical equations for $\kappa_1$ and $\kappa_2$ are shown in \cite{Note1}, including the case of $W=0$, for which $\kappa_2/\kappa_1 = 0$. The ratio $\kappa_2/\kappa_1$ allows to distinguish between Poisson level statistics, where $\sigma_{N}^{2} = m_N$ and $\kappa_2/\kappa_1 = 1$, and the statistics of a rigid spectrum, where neighboring eigenvalues strongly repel each other, yielding $\sigma_{N}^{2}=O(\ln N)$ and $\kappa_2/\kappa_1 = 0$ \cite{Altshuler88,Chalker1996,Bogomolny2011}. 

\begin{figure}[h]
\includegraphics[width=8cm,height=5.5cm]{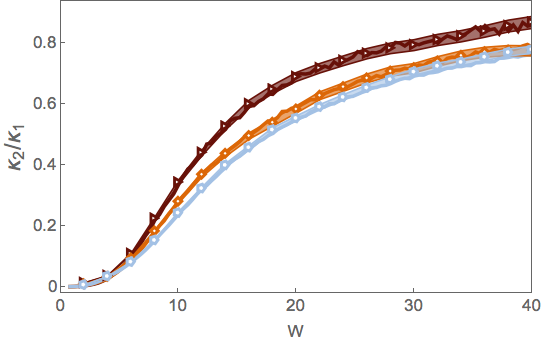}
\caption{Populations dynamics results (solid lines) for the cumulant ratio $\kappa_{2}/\kappa_1$  of the number of eigenvalues within $[-L,L]$ for the Anderson model on a regular random graph with connectivity $c=3$ and different $L$: $L=1/2$ (dark red),  $L=1$ (yellow) and $L=2$ (blue). We also present numerical diagonalization results (symbols) for matrices of size $N=1000$ and average over $5\times 10^3$ samples, for $L=1/2$ and $L = 1$, and over $10^4$ samples for $L=2$. The shaded area around each curve represents the error bars. }
\label{Andersonfig2}
\end{figure}

Figure \ref{Andersonfig2} displays population dynamics results for $\kappa_2/\kappa_1$ as a function of $W$ for fixed connectivity $c=3$. The 
level compressibility $\kappa_2/\kappa_1$ is a continuous and monotonic function of $W \geq 0$, which 
approaches $\kappa_2/\kappa_1 \rightarrow 1$ only for $W \rightarrow \infty$.
More interestingly, it fulfills $0 < \kappa_2/\kappa_1 < 1$ for any $W>0$ and, consequently, the number of energy levels inside $[-L,L]$
corresponding to extended eigenstates follows a sub-Poissonian statistics. This is in contrast to the behavior of the extended states in Gaussian random matrices \cite{Marino2014}, where $\kappa_2/\kappa_1 = 0$ for an interval of size $2L=O(1)$, due to the strong level-repulsion. The population dynamics results are free of finite size effects, as they arise from the solution of eqs. (\ref{ddfa}) and (\ref{eq:sce1}), valid for $N \rightarrow \infty$.

Finally, we remark that our results do not allow to draw conclusions on the existence of an ergodic/non-ergodic transition in the extended phase of the Anderson model on a RRG, since we 
have considered $L=O(1)$, independently of $N$. Such transition can be studied, in principle, by computing $\kappa_2/\kappa_1$ corresponding
to the statistics of low-lying energies $\{ \lambda_i \}$ that fulfill $1/N \ll \lambda_i \ll E_T$, where $E_T \propto (\ln N)^{-1}$ is the Thouless energy for the Anderson model on a RRG \cite{Mirlin}.
This is achieved by setting $L=s/N$, with $s \gg 1$ \cite{Mirlin}.
Although we do not study local eigenvalue fluctuations, our approach opens the very interesting
perspective that such problem can be addressed analytically by considering finite size corrections, following the ideas of \cite{MetzPar,Metz2015b}. Work along this line is underway.

\begin{acknowledgments}
The authors thank Konstantin Tikhonov, Alexander Mirlin and Mikhail Skvortsov for interesting comments. FLM 
thanks the hospitality of the Institute of Physics at UNAM.
This work has been funded by the program UNAM-DGAPA-PAPIIT IA101815. 
\end{acknowledgments}

\bibliography{biblio.bib}

\end{document}